\documentclass[12pt,a4paper]{article}
\usepackage{jheppub}
\usepackage{multirow}
\usepackage{chemarrow}
\usepackage{amsmath,amssymb,amsfonts,mathtools}
\usepackage{float}
\usepackage{subfig}
\usepackage{scalefnt}
\usepackage{wrapfig}
\usepackage{fancybox}
\usepackage{ifsym}
\usepackage{relsize}
\newcommand{\bea}{\begin{eqnarray}\displaystyle}
\newcommand{\eea}{\end{eqnarray}}

\newlength{\arrow}
{\setlength{\fboxsep}{15pt}
\setlength{\mylength}{\linewidth}%
\addtolength{\mylength}{-2\fboxsep}%
\addtolength{\mylength}{-2\fboxrule}%
\Sbox
\minipage{\mylength}%
\setlength{\abovedisplayskip}{0pt}%
\setlength{\belowdisplayskip}{0pt}%
\equation}%
{\endequation\endminipage\endSbox
\[\fbox{\TheSbox}\]}

\def\fint{\makebox[0pt][l]{\hspace{3.4pt}$-$}\int}
\def\CC{{\cal C}}
\def\CD{{\cal D}}
\def\CF{{\cal F}}
\def\CO{{\cal O}}
\def\CN{{\cal N}}

\begin{document}
\title{M-strings in thermodynamic limit: Seiberg-Witten geometry}
\author[\ast\ddagger]{Babak Haghighat,}
\author[\ast\dag]{Wenbin Yan}
\affiliation[\ast]{Jefferson Physical Laboratory, Harvard University, Cambridge, MA 02138, USA}
\affiliation[\ddagger]{Department of Mathematics, Harvard University, Cambridge, MA 02138, USA}
\affiliation[\dag]{Center of Mathematical Sciences and Applications, Harvard University, Cambridge, 02138, USA}

\abstract{
We initiate the study of M-strings in the thermodynamic limit. In this limit the BPS partition function of M5 branes localizes on configurations with a large number of strings which leads to a reformulation of the partition function in terms of a matrix model. We solve this matrix model and obtain its spectral curve which can be interpreted as the Seiberg-Witten curve associated to the compactified M5 brane theory.}
\maketitle

\section{Introduction}

Six-dimensional SCFTs have become one of the centers of focus in the study of superconformal field theories \cite{Heckman:2013pva,DelZotto:2014hpa,Ohmori:2014kda,DelZotto:2014fia,Heckman:2015bfa,DelZotto:2015isa,Heckman:2015ola,Cordova:2015vwa,Hayashi:2015fsa,Cordova:2015fha,Heckman:2015axa,Hayashi:2015zka,Hayashi:2015vhy,Heckman:2016ssk} and play an increasingly important role in the construction of four-dimensional quantum field theories \cite{DelZotto:2015rca,Morrison:2016nrt,Ohmori:2015pia,Ohmori:2015pua}.
M-strings \cite{Haghighat:2013gba} provide a framework for computing partition functions of six-dimensional SCFTs through elliptic genera of tensionless strings. The corresponding elliptic genera can be computed in various ways, either through a topological vertex or sigma model computation \cite{Haghighat:2013gba} or by using a quiver gauge theory description \cite{Haghighat:2013tka,Kim:2014dza,Haghighat:2014vxa,Gadde:2015tra,Kim:2015fxa}. In particular, for a theory that has $k$ different types of strings with tensions ($\varphi_1,\ldots, \varphi_k$) (which can be identified with scalar vevs in the associated tensor multiplets), one has for the 6d partition function on $T^2 \times \mathbb{R}^4$
\begin{equation}
	Z^{6d}_{\mathbb{R}^4\times T^2} = \sum_{\vec{n}} e^{-\vec{n} \cdot \vec{\varphi}} Z^{\vec{n}}_{T^2},
\end{equation}
where $Z^{\vec{n}}_{T^2}$ denotes the elliptic genus of $\vec{n}=(n_1,\cdots,n_k)$ strings.

In this note we will focus on the theory of M-strings which arise from M2-branes suspended between M5-branes \cite{Haghighat:2013gba} and analyse the behaviour of the partition function in the limit of large volume and large number of strings, the so called thermodynamic limit. In particular, the tensor deformation of the $(2,0)$ $A_1$ type theory giving rise to 5d $\mathcal{N}=1^*$ $SU(2)$ SYM is our main example. This theory has only one type of string with tension $\varphi$ which corresponds to the Coulomb branch vev of the five-dimensional gauge theory. Upon further compactification on a circle this vev becomes complexified and corresponds to the Seiberg-Witten period of the gauge theory. We will show how the Seiberg-Witten geometry of the torus compactification of the six-dimensional theory emerges in the thermodynamic limit of BPS partition functions and compute its differential. The methods used can be traced back to the work of Nekrasov and Okounkov on random partitions \cite{Nekrasov:2003rj} who used matrix model technology to compute the Seiberg-Witten geometry of four-dimensional $\mathcal{N}=2$ supersymmetric gauge theories. More precisely, important for our story will be the elliptic version of the results of Nekrasov and Okounkov first presented in \cite{Hollowood:2003cv}. However, our setup is considerably different from the one of \cite{Hollowood:2003cv} as we are using a string expansion of the partition function whereas the authors of \cite{Hollowood:2003cv} use an instanton expansion.

This paper is organized as follows. In section \ref{sec:m-strings} a review of the M-string setup will be presented. In section \ref{sec:matrixmodel} we will derive the matrix model description and from it the Seiberg-Witten geometry. Finally, in section \ref{sec:discussion}, we will present a discussion of results together with an outline for future work.

\section{Review of M-strings}
\label{sec:m-strings}

\begin{figure}[here!]
  \centering
	\includegraphics[width=0.5\textwidth]{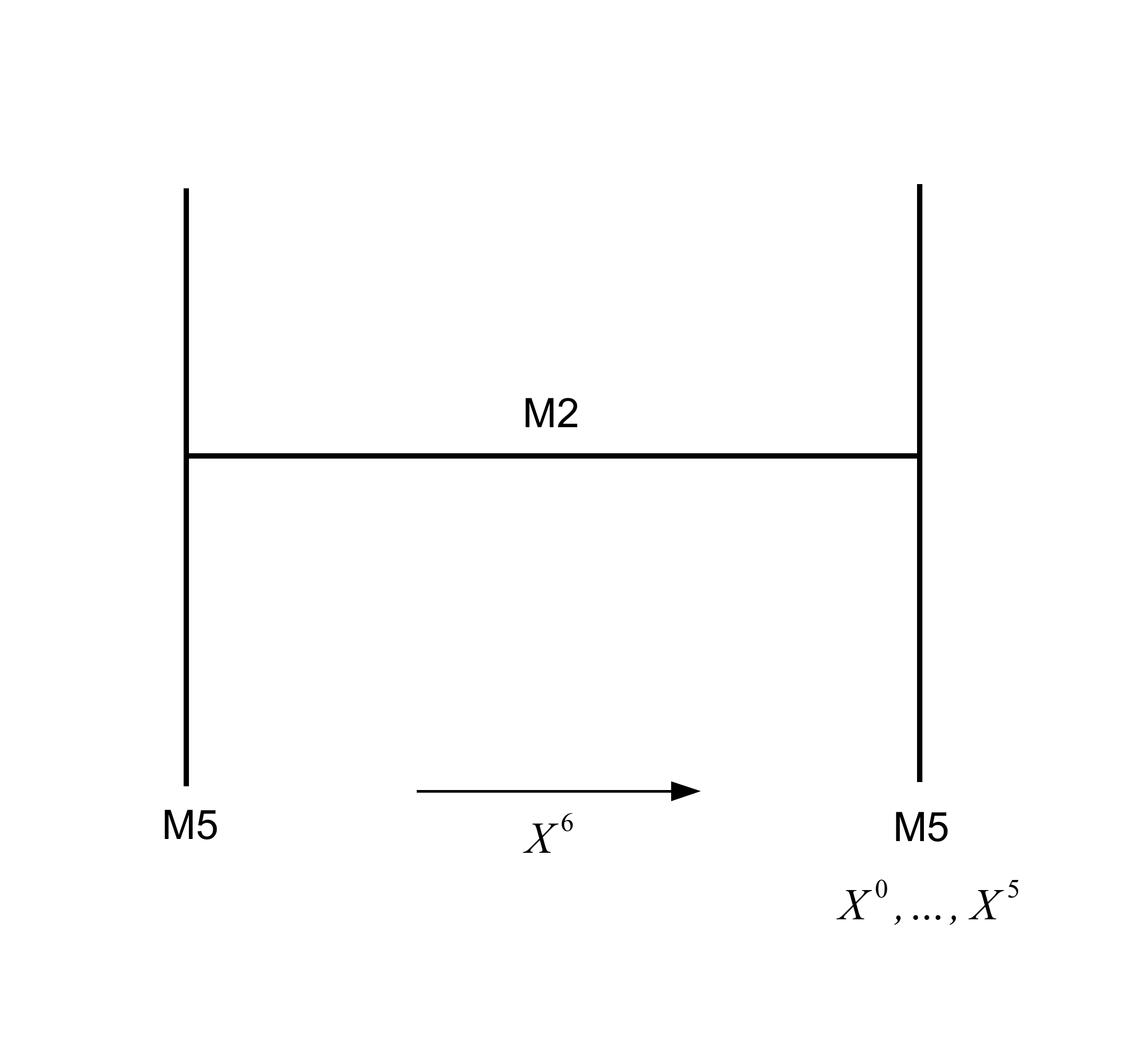}
  \caption{M2 branes suspended between M5 branes.}
  \label{fig:branesetup}
\end{figure}

M-strings arise from M2 branes suspended between M5 branes 
\cite{Haghighat:2013gba}. In the following we will present a review where we restrict ourselves to the case of two M5 branes which is the relevant situation for the present paper. We parametrize the eleven-dimensional spacetime of M-theory by coordinates $X^0, X^1, \ldots, X^{10}$, then consider M5 branes occupying the directions $X^0, X^1, X^2, X^3, X^4, X^5$ and separated along $X^6$. The worldvolume of each M5 brane is taken to be $T^2 \times \mathbb{R}^4$. For convenience, we parametrize $T^2$ by $X^0, X^1$ and $\mathbb{R}^4$ by $X^2, \ldots, X^5$. We extend this configuration by including M2 branes occupying the directions $X^0, X^1$ and $X^6$. These give rise to strings inside the M5 brane, which we denote by M-strings. This configuration is schematically shown in Figure \ref{fig:branesetup}.
The tension of strings is controlled by the distance between M5 branes, i.e. we have
\begin{equation}
	\langle X^6 \rangle = \varphi,
\end{equation}
where $\varphi$ is the scalar in the $(2,0)$ tensor multiplet. At low energies and for small $T^2$ this configuration gives rise to maximal Super Yang-Mills with gauge group $SU(2)$ in four dimensions and $\varphi$ becomes complexified and identified with the Coulomb branch vev of the gauge theory. The mass-deformation of maximal SYM can be incorporated by twisting the M5 brane geometry with a subgroup of the R-symmetry as explained in \cite{Haghighat:2013gba,Haghighat:2013tka}.

\begin{figure}[here!]
  \centering
	\includegraphics[width=0.8\textwidth]{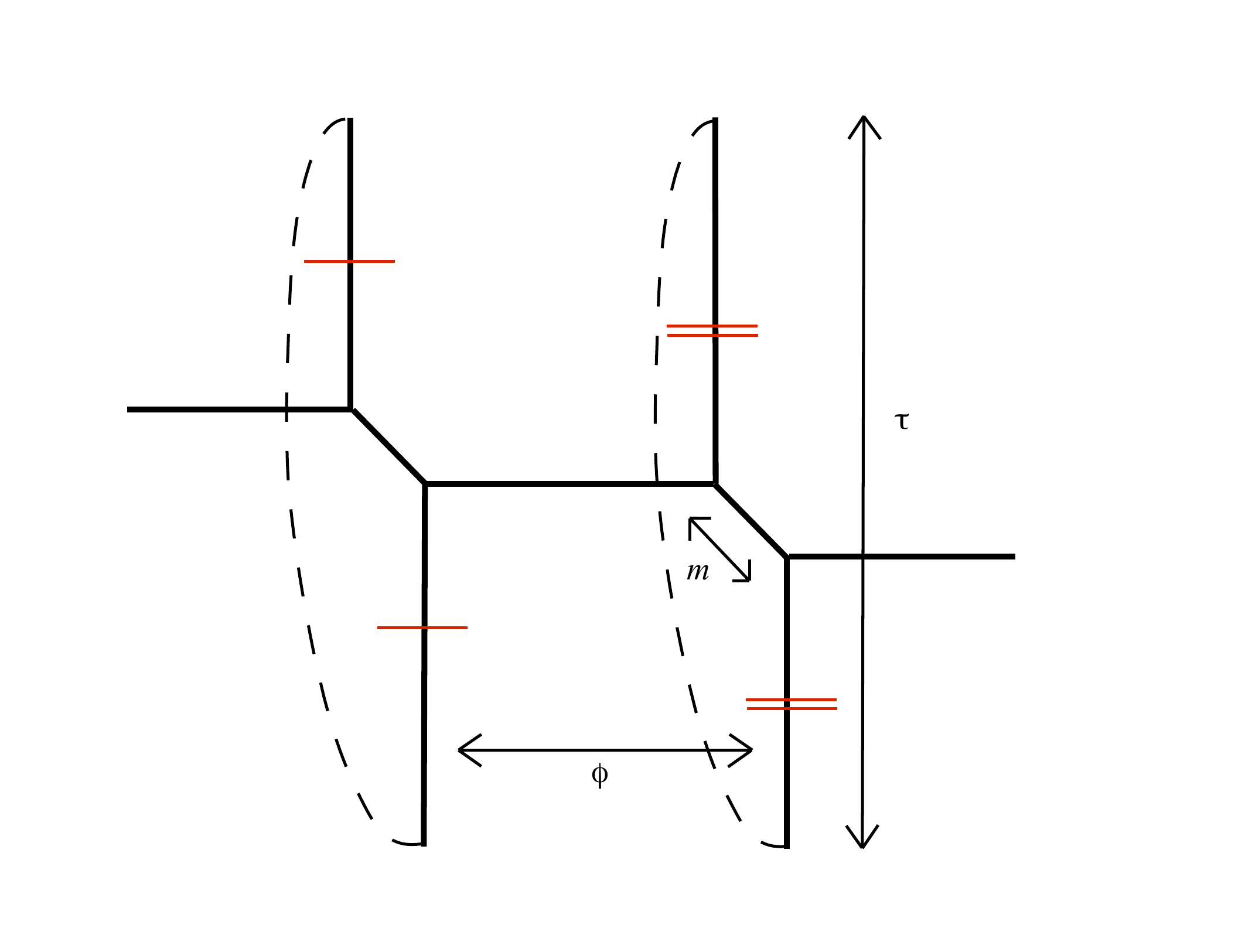}
  \caption{The toric geometry dual to the M5 brane configuration. $\tau$ is the complex structure parameter of the tours $T^2$ and $m$ denotes the mass of the adjoint hypermultiplet in the mass-deformation of maximal SYM. }
  \label{fig:toricg}
\end{figure}

Following a chain of dualities one can connect the M5 brane configuration just described to a toric non-compact Calabi-Yau geometry given by the toric skeleton graph depicted in Figure \ref{fig:toricg}.
In particular, the BPS partition function of the M5 brane configuration of Figure \ref{fig:branesetup} is equivalent to the topological string partition function of the toric Calabi-Yau of Figure \ref{fig:toricg}. 

\begin{figure}[here!]
  \centering
	\includegraphics[width=0.6\textwidth]{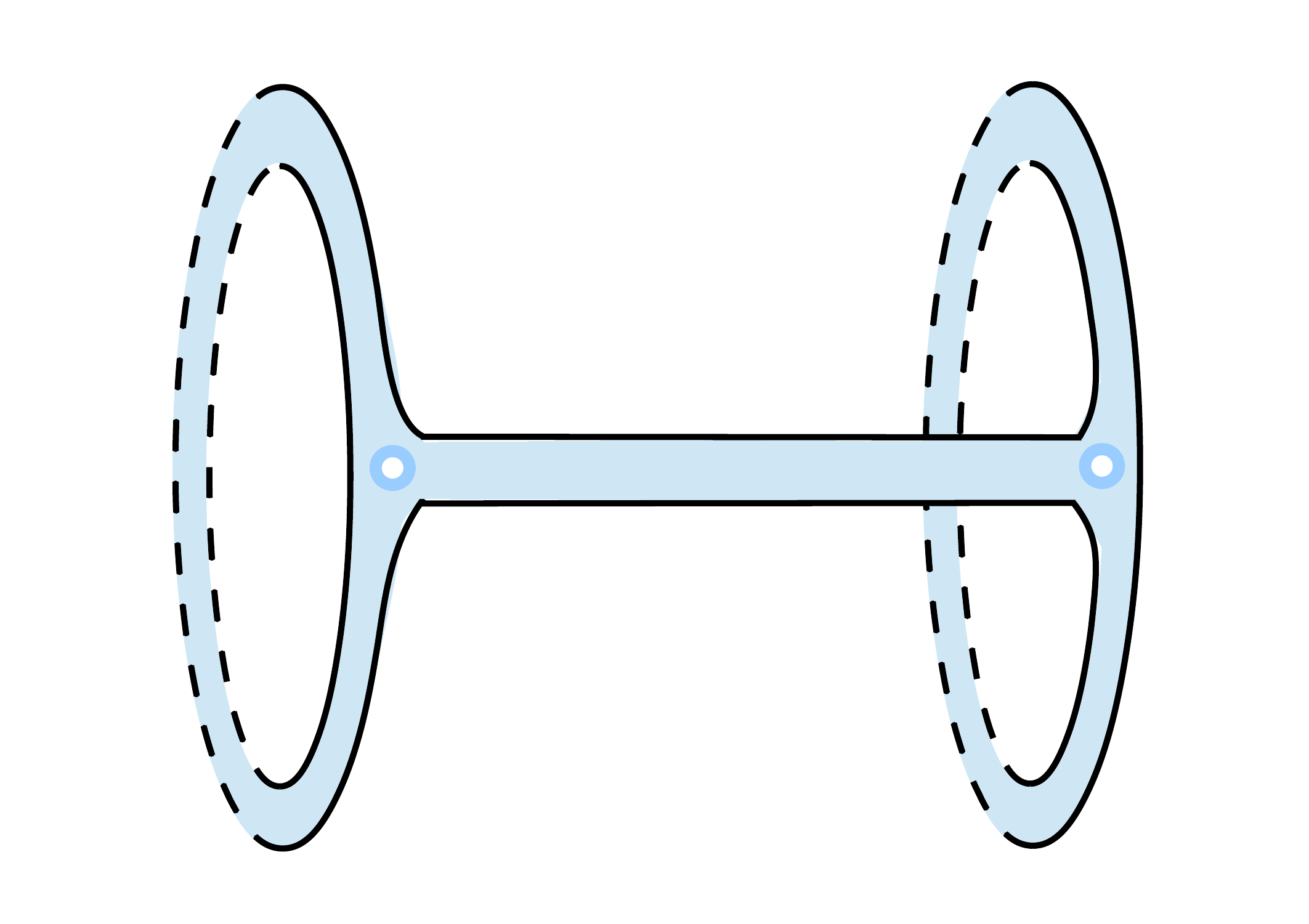}
  \caption{The Seiberg-Witten geometry arising from the thickening of the toric diagram. The punctures at the ends of the tube connecting the two tori correspond to the two outer legs of the toric diagram \ref{fig:toricg}.}
  \label{fig:SWg}
\end{figure}

 The mirror geometry and in particular the Seiberg-Witten curve associated to the above toric Calabi-Yau can be obtained as the ``thickening" of the toric diagram (see for example \cite{Aganagic:2001nx}). This is shown in Figure \ref{fig:SWg}.
In the next section we will show how this geometry naturally arises from the thermodynamic limit of M-strings. In order to proceed we will need the explicit form of the M5 brane partition function:
\begin{equation} \label{eq:MstrRes}
	Z^{\textrm{M5}} = Z^{\textrm{top}}=\sum_{N=1}^{\infty} e^{2\pi i \phi N} Z_N(\tau,\epsilon_1,\epsilon_2, m) = \sum_N e^{2\pi i \phi N} \sum_{|\nu|=N} \prod_{(i,j)\in \nu} \frac{\theta_1(\tau;z_{ij})\theta_1(\tau;v_{ij})}{\theta_1(\tau;w_{ij})\theta_1(\tau;u_{ij})},
\end{equation}
where the quantity $Z_N(\tau, \epsilon_1,\epsilon_2,m)$ is the elliptic genus of $N$ M-strings suspended between two M5 branes, $\nu$ is a Young tableau and $|\nu|$ denotes the number of boxes in $\nu$. The relation between the tension $\varphi$ of M2 branes and the parameter $\phi$ in the partition function (\ref{eq:MstrRes}) is
\begin{equation}
\varphi=-2\pi i \phi.
\end{equation}
Furthermore, $(i,j) \in \nu$ specifies a box in the $i$'th row and $j$'th column and the elliptic parameters as functions of $y = Q_m = e^{2\pi i m}$, $q=e^{2\pi i \epsilon_1}$ and $t = e^{-2\pi i \epsilon_2}$ are given by
\begin{equation}
\begin{array}{ll}
	e^{2\pi z_{ij}} = Q_m^{-1} q^{\nu_i - j + 1/2}t^{-i+1/2}, & e^{2\pi i v_{ij}} = Q_m^{-1} t^{i-1/2}q^{-\nu_i + j-1/2}, \nonumber \\
	e^{2\pi i w_{ij}} = q^{\nu_i -j +1} t^{\nu_j^t - i},  & e^{2\pi u_{ij}} = q^{\nu_i -j} t^{\nu_j^t - i +1},
	\end{array}
\end{equation}
and the theta function is defined as
\begin{equation}
	\theta_1(\tau;z) = i Q_{\tau}^{1/8} e^{\pi i z} \prod_{k=1}^{\infty}(1-Q_{\tau}^k)(1-Q_{\tau}^k e^{2\pi i z}) (1-Q_{\tau}^{k-1}e^{-2\pi i z}).
\end{equation}
For later convenience we note the following transformation property of the theta function under lattice shifts:
\begin{equation} \label{eq:thetatrf}
	\theta_1(\tau;z+n + l\tau) = e^{2\pi i (\frac{n-l}{2}-l z - \frac{l^2}{2}\tau)} \theta_1(\tau;z).
\end{equation}
 One crucial feature of the above partition function is that it is given in terms of an expansion in the ``Coulomb branch" parameter $e^{2\pi i \phi}$ whereas the usual Nekrasov partition function is always an expansion in terms of the ``instanton parameter" $e^{2\pi i\tau}$. The explanation of this unusual expansion is an underlying duality between the (mass-deformed) $A_1$ $(2,0)$ theory compactified on a circle and a 5d $\mathcal{N}=1$ $U(1)$ gauge theory with two fundamental hypermultiplets. For explanations and generalizations of this duality we refer to \cite{Haghighat:2013gba,  Haghighat:2013tka, DelZotto:2014hpa}. In this correspondence the parameter $\phi$ becomes the gauge coupling of the $U(1)$ theory:
\begin{equation} \nonumber
	\phi \longleftrightarrow \tau_{U(1)}~.
\end{equation}
This correspondence will serve as a guiding principle for the derivation of the Seiberg-Witten geometry which we will present in the next section.  For the remainder of this paper we will focus on the unrefined case
\begin{equation}
	\hbar = \epsilon_1 = -\epsilon_2,
\end{equation}
and leave the refined version for future study. In the following we will suppress the $\tau$-dependence of the theta function and simply write $\theta_1(z) \equiv \theta_1(\tau;z)$.

\section{Derivation of matrix model and Seiberg-Witten geometry}
\label{sec:matrixmodel}

In this section we take the thermodynamic limit of the partition function. This limit arises by considering the two limits\footnote{The $\hbar \rightarrow 0$ limit also can be thought of as the $V \rightarrow \infty$ limit. Here $V = \hbar^2$ is the regularized volume of $\mathbb{R}^4$ which can be deduced from the fact that the free energy of M-strings has the expansion $\mathcal{F} = \frac{\mathcal{F}_0}{\hbar^2} +  \mathcal{O}(\hbar^0)$ and the double-pole arises from the zero-modes of the strings along $\mathbb{R}^4$.}
\begin{equation}
	\hbar \rightarrow 0, \quad N \rightarrow \infty
\end{equation}
simultaneously, while keeping the product finite, i.e.
\begin{equation}
	\hbar^2 N = const.
\end{equation}
In this limit the partition function which is a sum over Young-tableaux becomes an integral over Young-tableaux profiles which can be evaluated by a saddle-point approximation. In the following we will provide a full self-consistent derivation of the resulting matrix model and its solution.

In the unrefined limit $\hbar=\epsilon_1=-\epsilon_2$, the partition function is
\begin{equation}
\label{eq:urpartition}
Z=\sum_{N}e^{2\pi i\phi N}
  \sum_{|\nu|=N}\prod_{(i,j)\in\nu}\frac{\theta_1(-m+\hbar(\nu_i-j-i+1))\theta_1(-m-\hbar(\nu_i-j-i+1))}{\theta_1(\hbar(\nu_i-j+\nu^\dag_j-i+1))^2}.
\end{equation}
Applying the identity
\begin{equation}
\prod_{i,j=1}^{\infty}\frac{\sigma(\nu_i-\nu_j+j-i)}{\sigma(j-i)}=\prod_{(i,j)\in\nu}\frac{1}{\sigma^2(\nu_i-j+\nu^\dag_j-i+1)},
\end{equation}
we rewrite (\ref{eq:urpartition}) as
\begin{equation}
\label{eq:urpartitionrewrite}
\begin{split}
Z=&\sum_\nu e^{2\pi i |\nu|\phi}Z_\nu,\\
Z_\nu=&\sum_{|\nu|}
  \prod_{i,j=1}^\infty \frac{\theta_1(\hbar(\nu_i-\nu_j+j-i))}{\theta_1(\hbar(j-i))}  \\
  &\times\prod_{(i,j)\in\nu}\theta_1(-m+\hbar(\nu_i-j-i+1))\theta_1(-m-\hbar(\nu_i-j-i+1)),
\end{split}
\end{equation}
here we take $\sigma(x)=\theta_1(\hbar x)$ since we are dealing with a 6d partition function.

We now take the thermodynamic limit of the partition function (\ref{eq:urpartitionrewrite}) and evaluate the prepotential in the similar way as in \cite{Nekrasov:2003rj,Hollowood:2003cv,Ishii:2013nba}. Firstly, we introduce a function $\gamma(z;\hbar)$ which obeys the difference equation,
\begin{equation}
\label{eq:difference}
\gamma(z+\hbar;\hbar)+\gamma(z-\hbar;\hbar)-2\gamma(z;\hbar)
=\ln\theta_1(z),
\end{equation}
and has the expansions,
\begin{equation}
\gamma(z;\hbar)=\sum_{g=0}^{\infty}h^{2g-2}\gamma_g(z).
\end{equation}
In the following derivation, we will use the fact that 
\begin{equation}
\gamma''_0(z)=\ln\theta_1(z).
\end{equation}
The explicit form of $\gamma(z;\hbar)$ is not important for us.

Using the difference equation (\ref{eq:difference}) the partition function (\ref{eq:urpartitionrewrite}) can be written as
\begin{equation}
\begin{split}
Z_\nu=\exp\bigg[&
-\frac{1}{4}\fint dz dw f''_\nu(z)f''_\nu(w)\gamma(z-w;\hbar)+\frac{1}{2}\fint dz f''_\nu(z)
\left(\gamma(z-m;\hbar)+\gamma(z+m;\hbar)\right)
\\
&+\gamma(0;\hbar)-\gamma(-m;\hbar)-\gamma(m;\hbar)\bigg],
\end{split}
\end{equation}
where we have defined the \textit{profile} $f_\nu(z)$ of the partition $\nu$ as follows
\begin{equation}
f_\nu(z)=\sum_{i=1}^{l(\nu)}\left(|z-\hbar(\nu_i-i+1)|-|z-\hbar(\nu_i-i)|\right)
+|z+\hbar l(\nu)|.
\end{equation}
Its second derivative is readily computed to be
\begin{equation}
f''_\nu(z)=
2\left[\sum_{i=1}^\infty\left(
\delta(z-\hbar(\nu_i-i+1))-\delta(z-\hbar(\nu_i-i))
-\delta(z+\hbar(i-1))+\delta(z+\hbar i)
\right)+\delta(z)\right].
\end{equation}
When $|\nu|$ is large, $f''_\nu(z)$ becomes a density function. Let $\CC$ be the support of $f''_\nu(z)$ around the orgin. The expression of $f''_\nu(z)$ leads to the constraints
\begin{equation}
\begin{split}
0=&\frac{1}{2}\int_\CC dz z f''_\nu(z),\\
|\nu|=&\frac{1}{4 \hbar^2}\int_\CC dz z^2 f''_\nu(z).
\end{split}
\end{equation}

In the thermodynamic limit $\hbar\rightarrow0$ the partition function is approximated by,
\begin{equation}
Z\simeq\int\CD f''d\lambda
\exp\left[\frac{1}{2\hbar^2}\CF_0+\CO(\hbar)\right],
\end{equation}
where
\begin{equation}
\begin{split}
\CF_0[f'',\lambda]
=&-\frac{1}{2}\fint_\CC dzdw f''(z)f''(w)\gamma_0(z-w)
+\fint_\CC dz f''(z)\left[\gamma_0(z-m)+\gamma_0(z+m)\right]\\
&+4\pi i\phi\left(\frac{1}{4}\int_\CC dz z^2 f''(z)\right)
+2\lambda\left(\frac{1}{2}\int_\CC dz z f''(z)\right).
\end{split}
\end{equation}
The saddle point equation is given by the variation of $\CF_0$ with respect to $f''$,
\begin{equation}
\label{eq:saddle}
\fint_\CC dw f''(w)\gamma_0(z-w)-\gamma_0(z-m)-\gamma_0(z+m)
-\pi i \phi z^2-\lambda z=0.
\end{equation}
Taking three derivatives with respect to $z$ we obtain
\begin{equation}
\omega(z+i\epsilon)+\omega(z-i\epsilon)=0,
\end{equation}
where $\omega$ is the resolvent,
\begin{equation}
\label{eq:resolvent}
\omega(z)=\fint_\CC f''(w)\partial_z\ln\theta_1(z-w)dw
-\partial_z\ln\theta_1(z-m)-\partial_z\ln\theta_1(z+m).
\end{equation}
The density function $f''$ is then the discontinuity of the resolvent $\omega$ along $\CC$,
\begin{equation}
2\pi i f''(z)=\omega(z-i\epsilon)-\omega(z+i\epsilon),\,\,\,\,\,z\in\CC.
\end{equation}

Equation (\ref{eq:resolvent}) shows that the resolvent $\omega$ is doubly periodic and has only simple poles at $\pm m$, therefore $\omega$ can be solved by the following ansatz,
\begin{equation}
\omega(z)=\frac{2\partial_z\sqrt{H(z)}}{\sqrt{H(z)-1}},
\end{equation}
where 
\begin{equation} \label{eq:Hfunction}
H(z)=c\frac{\theta_1(z)^2}{\theta_1(z-m)\theta_1(z+m)},
\end{equation}
where $c$ is a constant which we will fix later. $H(z)$ is chosen such that $\omega$ is doubly-periodic and its only singularities are simple poles at $\pm m$. These properties follow from the form (\ref{eq:resolvent}) of the resolvent. One can use the transformation property (\ref{eq:thetatrf}) to convince oneself of the fact that $H(z)$ is indeed doubly periodic. The ansatz (\ref{eq:Hfunction}) is motived from the duality to the $U(1)$ gauge theory with two fundamental hypermultiplets mentioned in section \ref{sec:m-strings} whose resolvent was worked out in \cite{Hollowood:2003cv}.

\begin{figure}[here!]
  \centering
	\includegraphics[width=0.6\textwidth]{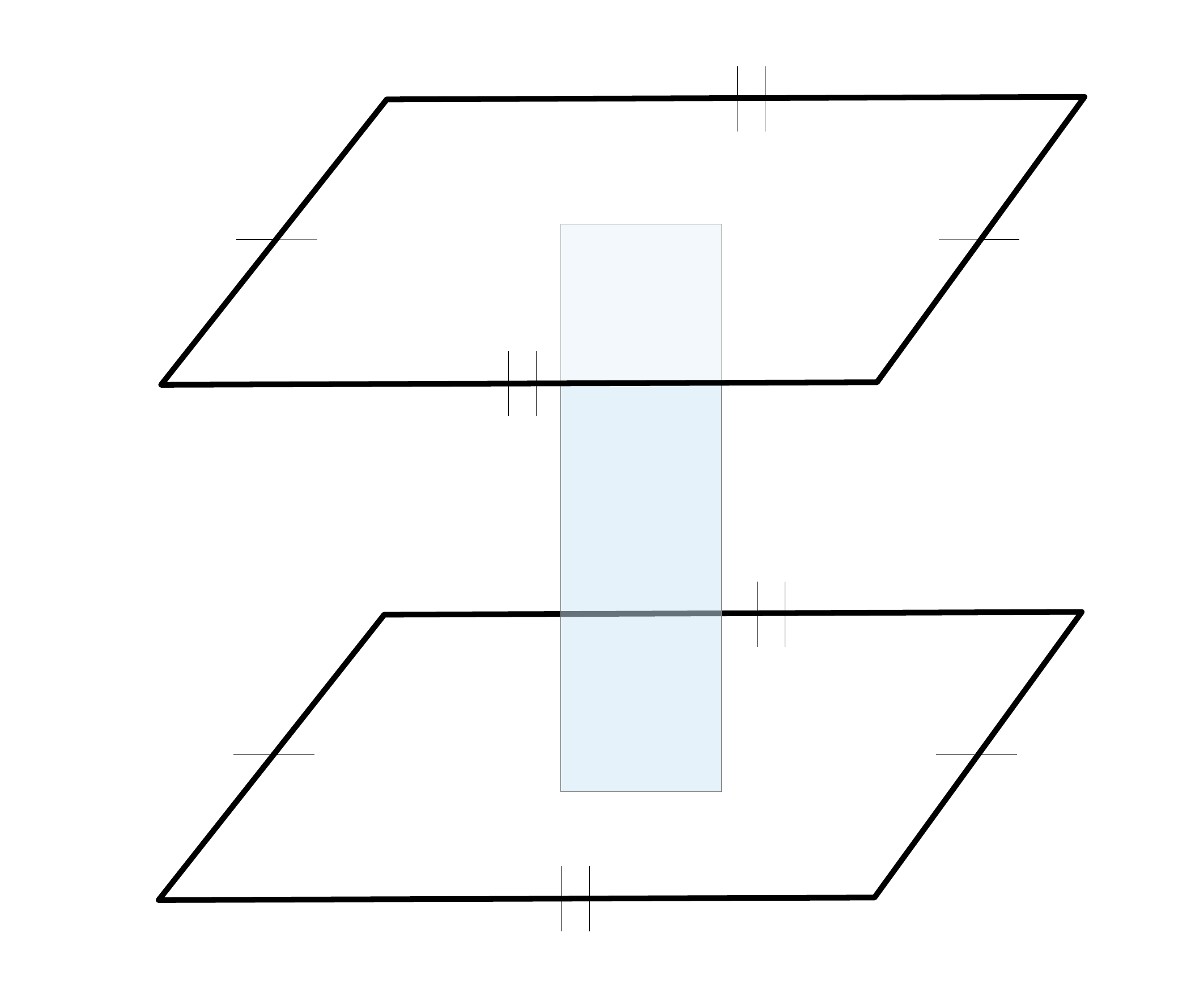}
  \caption{Seiberg-Witten geometry as double-cover of the torus.}
  \label{fig:doublecover}
\end{figure}

Now we are ready to work out the Seiberg-Witten geometry. Following the procedure in \cite{Hollowood:2003cv} the Seiberg-Witten curve can be written as
\begin{equation}
y^2=c\theta_1(z)^2-\theta_1(z-m)\theta_1(z+m).
\end{equation} To get the correct Seiberg-Witten period, we take two derivatives of the saddle point equation (\ref{eq:saddle}) giving
\begin{equation}
\frac{1}{2}\left(\Omega(z-i\epsilon)+\Omega(z+i\epsilon)\right)-2\pi i \phi=0,\,\,\,\,\,z\in\CC,
\end{equation}
where
\begin{equation}
\Omega(z)\equiv\int_\CC f''(w)\ln\theta_1(z-w)-\ln\theta_1(z-m)-\ln\theta_1(z+m),
\end{equation}
and $\Omega$ obeys
\begin{equation}
\Omega'(z)=\omega(z).
\end{equation}
Solving for $\phi$ we get
\begin{equation}
\phi=\frac{i}{2\pi^2}\oint_A\ln\left(\sqrt{H(z)}+\sqrt{H(z)-1}\right)dz\,\,\mathrm{mod}\,\mathbb{Z},
\end{equation}
where $A$ stands for the $A$-cycle on the torus. This reproduces the Seiberg-Witten geometry of two sheets connected through a branch cut. Here we observe that $\varphi = -2\pi i \phi$ is the Seiberg-Witten period of the circle compactification of the $\mathcal{N}=1^*$ $SU(2)$ gauge theory in five dimensions. Therefore, we deduce
\begin{equation}
	\varphi = \varphi(u,\tau)
\end{equation}
and compute
\begin{equation} \label{eq:SWperiod}
	\frac{\partial \varphi}{\partial u} = \frac{1}{2 \pi} \oint_A \frac{c'(u)}{c(u)} \frac{1}{\sqrt{1-\frac{1}{c}\frac{\theta_1(z)^2}{\theta_1(z-m)\theta_1(z+m)}}} dz.
\end{equation}
Similarly for $\varphi_D$ we have
\begin{equation}
	\frac{\partial \varphi_D}{\partial u} = \frac{1}{2 \pi} \oint_B \frac{c'(u)}{c(u)} \frac{1}{\sqrt{1-\frac{1}{c}\frac{\theta_1(z)^2}{\theta_1(z-m)\theta_1(z+m)}}} dz.
\end{equation}
To fix the constant $c$ we take the massless limit,
\begin{equation}
\begin{split}
\lim_{m\rightarrow0}\frac{\partial\varphi}{\partial u}
=\frac{1}{2\pi}\oint_A\frac{c'(u)}{\sqrt{c}}\frac{1}{\sqrt{c-1}}dz,\\
\lim_{m\rightarrow0}\frac{\partial\varphi_D}{\partial u}
=\frac{1}{2\pi}\oint_B\frac{c'(u)}{\sqrt{c}}\frac{1}{\sqrt{c-1}}dz.
\end{split}
\end{equation}
By matching this to the well-know Seiberg-Witten theory of massless $\CN=2^{\star}$ \cite{Seiberg:1994aj}, 
\begin{equation}
\left(\frac{\partial\varphi}{\partial u},\frac{\partial\varphi_D}{\partial u}\right)
=\frac{\sqrt{2/u}}{u}(1,\tau),
\end{equation}
we deduce
\begin{equation}
c=\cosh^2(\sqrt{u}),
\end{equation}
where we have made us of the fact that
\begin{equation}
	\oint_A dz = 2\pi, \quad \oint_B dz = 2\pi\tau.
\end{equation}
Let us next comment on the emerging Seiberg-Witten geometry. We observe that for finite values of $u$ this geometry is given by two sheets which are doubly-periodic and connected through one branch cut as shown in Figure \ref{fig:doublecover}.
For large values of $u$ the function $c(u)$ approaches infinity and therefore from the form (\ref{eq:SWperiod}) we see that the branch cut disappears.

\section{Discussion}
\label{sec:discussion}

In this note we study the Seiberg-Witten geometry of the simplest M-string configuration via the thermodynamic limit of the BPS partition function of M5 branes. The resulting geometry being two sheets connected by a branch cut matches exactly the prediction from the toric geometry. It would be interesting to extend these results in the following directions. A first direction is to apply our methods to the case of ADE configurations of 6d String Chains arising from D5 branes probing ADE singularities \cite{Gadde:2015tra} and extract the resulting Seiberg-Witten geometries. This should be straight-forward as all string contributions can be written fully in terms of summations over Young diagrams. Another path to follow is the study of the NS-deformation of the SW-curve, the so called quantum curve as explored for example in \cite{Huang:2012kn}. Finally, an important extension of the present results would be the derivation of the thermodynamic limit of strings which are in the class of minimal 6d SCFT's as studied in \cite{Haghighat:2014vxa}. Here, there is no closed Young diagram representation available and therefore the derivation of the saddle point equation more challenging. 

\section*{Acknowledgment}

We would like to thank M. Aganagic, S. Cherkis, J. Gu, A. Klemm, C. Kozcaz for helpful discussions. BH would like to thank the Berkeley Center for Theoretical Physics for hospitality where part of this work was completed. The work of WY is supported by the Center for Mathematical Sciences and Applications at Harvard University.

\bibliography{references}

\begin{thebibliography}{99}

\bibitem{Heckman:2013pva} 
  J.~J.~Heckman, D.~R.~Morrison and C.~Vafa,
  ``On the Classification of 6D SCFTs and Generalized ADE Orbifolds,''
  JHEP {\bf 1405}, 028 (2014)
  Erratum: [JHEP {\bf 1506}, 017 (2015)]
  doi:10.1007/JHEP06(2015)017, 10.1007/JHEP05(2014)028
  [arXiv:1312.5746 [hep-th]].
  
\bibitem{DelZotto:2014hpa} 
  M.~Del Zotto, J.~J.~Heckman, A.~Tomasiello and C.~Vafa,
  ``6d Conformal Matter,''
  JHEP {\bf 1502}, 054 (2015)
  doi:10.1007/JHEP02(2015)054
  [arXiv:1407.6359 [hep-th]].
  
\bibitem{Ohmori:2014kda} 
  K.~Ohmori, H.~Shimizu, Y.~Tachikawa and K.~Yonekura,
  ``Anomaly polynomial of general 6d SCFTs,''
  PTEP {\bf 2014}, no. 10, 103B07 (2014)
  doi:10.1093/ptep/ptu140
  [arXiv:1408.5572 [hep-th]].
  
\bibitem{DelZotto:2014fia} 
  M.~Del Zotto, J.~J.~Heckman, D.~R.~Morrison and D.~S.~Park,
  ``6D SCFTs and Gravity,''
  JHEP {\bf 1506}, 158 (2015)
  doi:10.1007/JHEP06(2015)158
  [arXiv:1412.6526 [hep-th]].
  
\bibitem{Heckman:2015bfa} 
  J.~J.~Heckman, D.~R.~Morrison, T.~Rudelius and C.~Vafa,
  ``Atomic Classification of 6D SCFTs,''
  Fortsch.\ Phys.\  {\bf 63}, 468 (2015)
  doi:10.1002/prop.201500024
  [arXiv:1502.05405 [hep-th]].
  
\bibitem{DelZotto:2015isa} 
  M.~Del Zotto, J.~J.~Heckman, D.~S.~Park and T.~Rudelius,
  ``On the Defect Group of a 6D SCFT,''
  Lett.\ Math.\ Phys.\  {\bf 106}, no. 6, 765 (2016)
  doi:10.1007/s11005-016-0839-5
  [arXiv:1503.04806 [hep-th]].
  
\bibitem{Heckman:2015ola} 
  J.~J.~Heckman, D.~R.~Morrison, T.~Rudelius and C.~Vafa,
  ``Geometry of 6D RG Flows,''
  JHEP {\bf 1509}, 052 (2015)
  doi:10.1007/JHEP09(2015)052
  [arXiv:1505.00009 [hep-th]].
  
\bibitem{Cordova:2015vwa} 
  C.~Cordova, T.~T.~Dumitrescu and X.~Yin,
  ``Higher Derivative Terms, Toroidal Compactification, and Weyl Anomalies in Six-Dimensional (2,0) Theories,''
  arXiv:1505.03850 [hep-th].
  
\bibitem{Hayashi:2015fsa} 
  H.~Hayashi, S.~S.~Kim, K.~Lee, M.~Taki and F.~Yagi,
  ``A new 5d description of 6d D-type minimal conformal matter,''
  JHEP {\bf 1508}, 097 (2015)
  doi:10.1007/JHEP08(2015)097
  [arXiv:1505.04439 [hep-th]].
  
\bibitem{Cordova:2015fha} 
  C.~Cordova, T.~T.~Dumitrescu and K.~Intriligator,
  ``Anomalies, Renormalization Group Flows, and the a-Theorem in Six-Dimensional (1,0) Theories,''
  arXiv:1506.03807 [hep-th].
 
  
\bibitem{Heckman:2015axa} 
  J.~J.~Heckman and T.~Rudelius,
  ``Evidence for C-theorems in 6D SCFTs,''
  JHEP {\bf 1509}, 218 (2015)
  doi:10.1007/JHEP09(2015)218
  [arXiv:1506.06753 [hep-th]].
 
  
\bibitem{Hayashi:2015zka} 
  H.~Hayashi, S.~S.~Kim, K.~Lee and F.~Yagi,
  ``6d SCFTs, 5d Dualities and Tao Web Diagrams,''
  arXiv:1509.03300 [hep-th].
  
\bibitem{Hayashi:2015vhy} 
  H.~Hayashi, S.~S.~Kim, K.~Lee, M.~Taki and F.~Yagi,
  ``More on 5d descriptions of 6d SCFTs,''
  arXiv:1512.08239 [hep-th].  
  
\bibitem{Heckman:2016ssk} 
  J.~J.~Heckman, T.~Rudelius and A.~Tomasiello,
  ``6D RG Flows and Nilpotent Hierarchies,''
  arXiv:1601.04078 [hep-th].
  
\bibitem{DelZotto:2015rca} 
  M.~Del Zotto, C.~Vafa and D.~Xie,
  ``Geometric engineering, mirror symmetry and $ 6{\mathrm{d}}_{\left(1,0\right)}\to 4{\mathrm{d}}_{\left(\mathcal{N}=2\right)} $,''
  JHEP {\bf 1511}, 123 (2015)
  doi:10.1007/JHEP11(2015)123
  
\bibitem{Morrison:2016nrt} 
  D.~R.~Morrison and C.~Vafa,
  ``F-Theory and N=1 SCFTs in Four Dimensions,''
  arXiv:1604.03560 [hep-th].
  
\bibitem{Ohmori:2015pia} 
  K.~Ohmori, H.~Shimizu, Y.~Tachikawa and K.~Yonekura,
  ``6d $\mathcal{N}=\left(1,\;0\right) $ theories on S$^{1}$ /T$^{2}$ and class S theories: part II,''
  JHEP {\bf 1512}, 131 (2015)
  doi:10.1007/JHEP12(2015)131
  [arXiv:1508.00915 [hep-th]].
  
\bibitem{Ohmori:2015pua} 
  K.~Ohmori, H.~Shimizu, Y.~Tachikawa and K.~Yonekura,
  ``6d $\mathcal{N}=(1,0)$ theories on $T^2$ and class S theories: Part I,''
  JHEP {\bf 1507}, 014 (2015)
  doi:10.1007/JHEP07(2015)014
  [arXiv:1503.06217 [hep-th]].
  
\bibitem{Haghighat:2013gba} 
  B.~Haghighat, A.~Iqbal, C.~Kozçaz, G.~Lockhart and C.~Vafa,
  ``M-Strings,''
  Commun.\ Math.\ Phys.\  {\bf 334}, no. 2, 779 (2015)
  doi:10.1007/s00220-014-2139-1
  [arXiv:1305.6322 [hep-th]].

\bibitem{Haghighat:2013tka} 
  B.~Haghighat, C.~Kozcaz, G.~Lockhart and C.~Vafa,
  ``Orbifolds of M-strings,''
  Phys.\ Rev.\ D {\bf 89}, no. 4, 046003 (2014)
  doi:10.1103/PhysRevD.89.046003
  [arXiv:1310.1185 [hep-th]].
  
\bibitem{Kim:2014dza} 
  J.~Kim, S.~Kim, K.~Lee, J.~Park and C.~Vafa,
  ``Elliptic Genus of E-strings,''
  arXiv:1411.2324 [hep-th].
  
\bibitem{Haghighat:2014vxa} 
  B.~Haghighat, A.~Klemm, G.~Lockhart and C.~Vafa,
  ``Strings of Minimal 6d SCFTs,''
  Fortsch.\ Phys.\  {\bf 63}, 294 (2015)
  doi:10.1002/prop.201500014
  [arXiv:1412.3152 [hep-th]].
  
\bibitem{Gadde:2015tra} 
  A.~Gadde, B.~Haghighat, J.~Kim, S.~Kim, G.~Lockhart and C.~Vafa,
  ``6d String Chains,''
  arXiv:1504.04614 [hep-th].
  
\bibitem{Kim:2015fxa} 
  J.~Kim, S.~Kim and K.~Lee,
  ``Higgsing towards E-strings,''
  arXiv:1510.03128 [hep-th].
  
\bibitem{Nekrasov:2003rj} 
  N.~Nekrasov and A.~Okounkov,
  ``Seiberg-Witten theory and random partitions,''
  Prog.\ Math.\  {\bf 244}, 525 (2006)
  [hep-th/0306238].
  
\bibitem{Hollowood:2003cv} 
  T.~J.~Hollowood, A.~Iqbal and C.~Vafa,
  ``Matrix models, geometric engineering and elliptic genera,''
  JHEP {\bf 0803}, 069 (2008)
  doi:10.1088/1126-6708/2008/03/069
  [hep-th/0310272].
  
\bibitem{Aganagic:2001nx} 
  M.~Aganagic, A.~Klemm and C.~Vafa,
  ``Disk instantons, mirror symmetry and the duality web,''
  Z.\ Naturforsch.\ A {\bf 57}, 1 (2002)
  doi:10.1515/zna-2002-1-201
  [hep-th/0105045].

 
\bibitem{Ishii:2013nba} 
  T.~Ishii and K.~Sakai,
  ``Thermodynamic limit of the Nekrasov-type formula for E-string theory,''
  JHEP {\bf 1402}, 087 (2014)
  doi:10.1007/JHEP02(2014)087
  [arXiv:1312.1050 [hep-th]].
  
\bibitem{Seiberg:1994aj} 
  N.~Seiberg and E.~Witten,
  ``Monopoles, duality and chiral symmetry breaking in N=2 supersymmetric QCD,''
  Nucl.\ Phys.\ B {\bf 431}, 484 (1994)
  doi:10.1016/0550-3213(94)90214-3
  [hep-th/9408099].
  
\bibitem{Huang:2012kn} 
  M.~x.~Huang,
  ``On Gauge Theory and Topological String in Nekrasov-Shatashvili Limit,''
  JHEP {\bf 1206}, 152 (2012)
  doi:10.1007/JHEP06(2012)152
  [arXiv:1205.3652 [hep-th]].


\end{thebibliography}

\end{document}